\documentclass[journal]{IEEEtran}
\ifCLASSINFOpdf
\else
\fi
\usepackage{amsmath}
\usepackage{lipsum}
\usepackage{graphicx}
\usepackage[justification=centering]{caption}
\usepackage{epstopdf}
\usepackage{tabularx}
\usepackage{algorithm}
\usepackage{algorithmic}
\usepackage{amssymb}

\usepackage{xcolor}

\begin{document}
%
\title{VeSoNet: Traffic-Aware Content Caching for Vehicular Social Networks based on Path Planning and Deep Reinforcement Learning }
%
%
%

\author{Nyothiri Aung, Sahraoui Dhelim, Liming Chen, Wenyin Zhang, Abderrahmane Lakas and Huansheng Ning
	\thanks{Nyothiri Aung, Sahraoui Dhelim and Wenyin Zhang are with the School of Information Science and Engineering, Linyi University, Shandong 276000, China} 
	\thanks{Liming Chen is with the School of Computing, Ulster University, Newtownabbey BT37 0QB, U.K.}
	\thanks{Abderrahmane Lakas is with the College of Information Technology (CIT), United Arab Emirates University, Al Ain, UAE}
	\thanks{Huansheng Ning	is with the School of Computer and Communication Engineering, University of Science and Technology Beijing, 100083, Beijing, China.}
	\thanks{Corresponding author: Huansheng Ning (ninghuansheng@ustb.edu.cn).}
}

%
%

\markboth{Journal of \LaTeX\ Class Files,~Vol.~14, No.~8, August~2015}%
{Shell \MakeLowercase{\textit{et al.}}: Bare Demo of IEEEtran.cls for IEEE Journals}
%



\maketitle

\begin{abstract}
Vehicular social networking is an emerging application of the promising Internet of Vehicles (IoV) which aims to achieve the seamless integration of vehicular networks and social networks. However, the unique characteristics of vehicular networks such as high mobility and frequent communication interruptions make content delivery to end-users under strict delay constrains an extremely challenging task. In this paper, we propose a social-aware vehicular edge computing architecture that solves the content delivery problem by using some of the vehicles in the network as edge servers that can store and stream popular content to close-by end-users. The proposed architecture includes three components. First, we propose a social-aware graph pruning search algorithm that computes and assigns the vehicles to the shortest path with the most relevant vehicular content providers. Secondly, we use a traffic-aware content recommendation scheme to recommend relevant content according to their social context. This scheme uses graph embeddings in which the vehicles are represented by a set of low-dimension vectors (vehicle2vec) to store information about previously consumed content. Finally, we propose a Deep Reinforcement Learning (DRL) method to optimize the content provider vehicles distribution across the network. The results obtained from a realistic traffic simulation show the effectiveness and robustness of the proposed system when compared to the state-of-the-art baselines.
\end{abstract}


%
\IEEEpeerreviewmaketitle

\section{Introduction}
%
%
%
%
\IEEEPARstart{W}{ith} the emergence of Internet of Vehicles (IoV) as a new networking paradigm that interconnect the vehicles with the ubiquitous Internet of Things (IoT) network, and the increasing adoption of 5G network in many countries, the vision of Intelligent Transportation System (ITS) is closer to realization than ever. The IoV network is expected to enhance many application domains and offer a plethora of smart services, ranging from essential emergence services to entertainment applications. Currently there are more than 1.4 billion vehicles worldwide, and this figure is expected to reach 3 billion in 2037 \cite{Voelcker}, which will worsen the existing traffic congestion problem. As more and more people spend hours in traffic congestion, they turn to social media and other entertainment services to spend the waiting time \cite{Aung2020}. The IoV can offer an alternative to connect the users with the Internet, and seamlessly interconnect their existing social networks to a vehicular social networking model, that brings social content near to passengers and reduce the expensive access to 4G/5G network. 

One of the most challenging problems in such vehicular social networking model is how to enable passengers to seamlessly access social network content without interruptions and content delivery delay. In vehicular networks, the content can be delivered through Vehicle to Infrastructure (V2I) communication with the Roadside Units (RSU) that is connected to the Internet, or through cellular base station using 4G/5G interface. The former method has the advantage of low costs and convenient communication but suffers from the drawback of difficult access, and a vehicle also needs to rely on Vehicle-to-Vehicle (V2V) communication to reach the sparse RSUs. While the latter has the advantage of wide coverage and instance access, it has the drawback of expensive communication \cite{Zhang2018}. The intuitive approach is to store the content of social network on a centralized server on the cloud, and the vehicles can access it through V2I communication or by downloading it using 4G/5G cellular network. Nonetheless, V2I communication is not suitable for live streaming due to the high speed of vehicles and the frequent disconnections between vehicles and RSU \cite{Dhelim2021}. On the other hand, the vehicle to base station communication is more stable compared to V2I communication, but not suitable for large file download due to high cost of network usage \cite{Abdenacer2021}. 

Motivated by the above-mentioned limitations, in this paper, we propose traffic-aware VEhicular SOcial NETwork (VeSoNet) content caching architecture that leverages vehicular edge computing paradigm to store and distribute the social content in vehicles, and bring the most popular content nearby end-users and cache it in the vehicles for future access. Our contributions can be summarized as follow:

\begin{itemize}
	\item We proposed a social-aware hybrid content distribution scheme, where only the popular data content is replicated and stored in data provider vehicles to minimize  download time, and the content provider vehicles follow vehicular edge computing paradigm to deliver the requested content.
	
	\item We develop a social-aware graph pruning search algorithm that computes and assigns the content consumer vehicles to the shortest path with the most relevant content providers, and leverage a Deep Reinforcement Learning DRL model to optimize  content provider vehicles distribution across the network.
	
	\item We develop a traffic-aware content recommendation approach based on graph embeddings named vehicle2vec, where vehicles are represented by a set of low dimensional vectors of their previously consumed content.
\end{itemize}

The rest of the paper is organized as follows: In Section 2, we review the literature of content caching and content delivery using vehicular edge computing, as well as social vehicular networks. In Section 3, we present the main components of the proposed vehicular network architecture. Section 4 details the system modeling of the proposed system. While in Section 5, we present the experimental evaluation and discuss the obtained results. Finally, we conclude the paper in Section 6.

\section{Related work}

Many previous studies have proposed different architectural designs for content delivery in vehicular networks. Dzyiauddin \textit{et al.} \cite{dziyauddin2021} surveyed computational offloading and content delivery and caching in vehicular edge computing, including the architectures, communication layers and applications of vehicular edge computing for content delivery and content caching. Zhang \textit{et al.} \cite{zhang2019deep} introduced a social-aware mobile edge computing architecture for content caching, where they employed DRL model, their proposed method takes advantage of the relationships among vehicles and RSUs to perform content dissemination with diverse vehicular social characteristics for urban environments. Whereas in other work \cite{zhang2021digital}, they extended their model by introducing digital twin technology to map the edge caching system into cyberspace, and used vehicular cloud to coordinate the correlation of the cached content among multiple vehicles, then employed deep-learning based route selection method that considers the social context, the vehicular cloud formation and cache resource allocation. Similarly, Qiao \textit{et al.} \cite{qiao2019deep} introduced a cooperative vehicular edge caching system to jointly optimize the content delivery and content placement  in the vehicular edge computing environment, with the aid of flexible cooperation between cellular stations, RSUs, and vehicular nodes. In this system, the joint optimization problem is modelled as a double time-scale Markov decision process (DTS-MDP). Zhou \textit{et al.} \cite{Zhou2019} introduced a new content delivery architecture by utilizing the 5G edge networks, where the content caching and data pre-fetching methods are discussed. They furthermore studied the comprehensive dynamic link utilization problem in 5G edge networks from the perspectives of network operator, as well as vehicle users. Luo \textit{et al.} \cite{Luo2020} introduced EdgeVCD, an intelligent algorithm-inspired content distribution mechanism, which uses a dual-importance (DI) evaluation method to reflect the relationship between the Priority of Vehicles (PoV) and the Priority of Contents (PoC), and formulate an optimization problem to maximize the system utility for content distribution

De Souza \textit{et al.} \cite{de2019safe} proposed Safe and Sound (SNS), which uses a hybrid architecture and a cooperative re-routing method to enhance the system computation performance and scalability. SNS utilizes a recurrent neural network (RRN) to predict future safety risks dynamics, as well as to offer a customized re-routing in which every vehicle choose the risks to avoid.  When the traffic server detects a congestion road, it notifies the incoming vehicles by sending the traffic report to all vehicles that their path cross through this road segment. The purpose of SNS's re-routing strategy is to balance the traffic flow over a set of alternative paths for every vehicle based on their current and final positions, as well as their preferences. Soua \textit{et al.} \cite{soua2017sdn} proposed a vehicular social networking architecture that combines content centric networking (CCN) model, Floating Content (CF), and Software Defined Networking (SDN) to offer a multi-pronged approach for adaptive content delivery. Where the interest in a given data content includes the location and name of the content requester. Intermediate nodes that receive the content interest message check their local content store (CS), and in case the requested content is not available in CS, they forward the message and trigger a timer. When the interest timer is expired, they retransmit the content interest message and update their pending interest table (PIT) accordingly. While the FC is used to support geographic content routing. A SDN controller is leveraged to operate a direct path between content requester and content provider, which is similar to the dynamic unicast method.  Alowish \textit{et al.} \cite{alowish2020novel} proposed content delivery architecture of vehicular networks named Cuckoo, in which the content is delivered by the RSU with the help of controller nodes. To deliver the content the RSU selects the optimal route by fetching the location of the data provider through the controller node. The optimal route selection is computed using the cuckoo search algorithm. Similarly, Zhao et al. \cite{zhao2007data} proposed DP-IB, a  vehicular content delivery system that uses a data pouring and buffering mechanism for content dissemination in VANET, where data contents are sent by the source node are buffered along the way and rebroadcasted in the road intersections. The main idea behind Data Pouring (DP) is instead of broadcasting the content data to the whole network, the scheme only sends the disseminated content to few road segments known as \textit{axis roads (A-Roads)}. The A-Roads are chosen as the main roads where the data center is located, and these road segments usually have denser traffic flow than other roads. While the Intersection Buffering (IB) means that the scheme also disseminate the content to nodes traveling along the crossing roads (C-Roads) that intersect with the A-Roads.

To deal with the privacy leakage problem in social vehicular networks, Zhang et al. \cite{zhang2020decentralized} introduced a distributed location privacy-preserving spatial crowdsourcing method for IoV, which enable vehicular nodes to take part of the spatial crowdsourcing and guarantee the privacy of task's location information. They employed blockchain to record the user data without the need for a centralized spatial crowdsourcing server. While Kang et al. \cite{kang2018blockchain} investigated the usage of blockchain and smart contract to improve the data storage security during content sharing among vehicles in vehicular edge networks. They concluded that blockchain technology can help achieving content sharing without authorization, furthermore, they introduced a reputation-based content sharing scheme to guarantee the data quality shared among vehicular nodes.

\section{Vehicular edge architecture}

\begin{figure*}[!htbp]
	\centering
	\includegraphics[width=\textwidth]{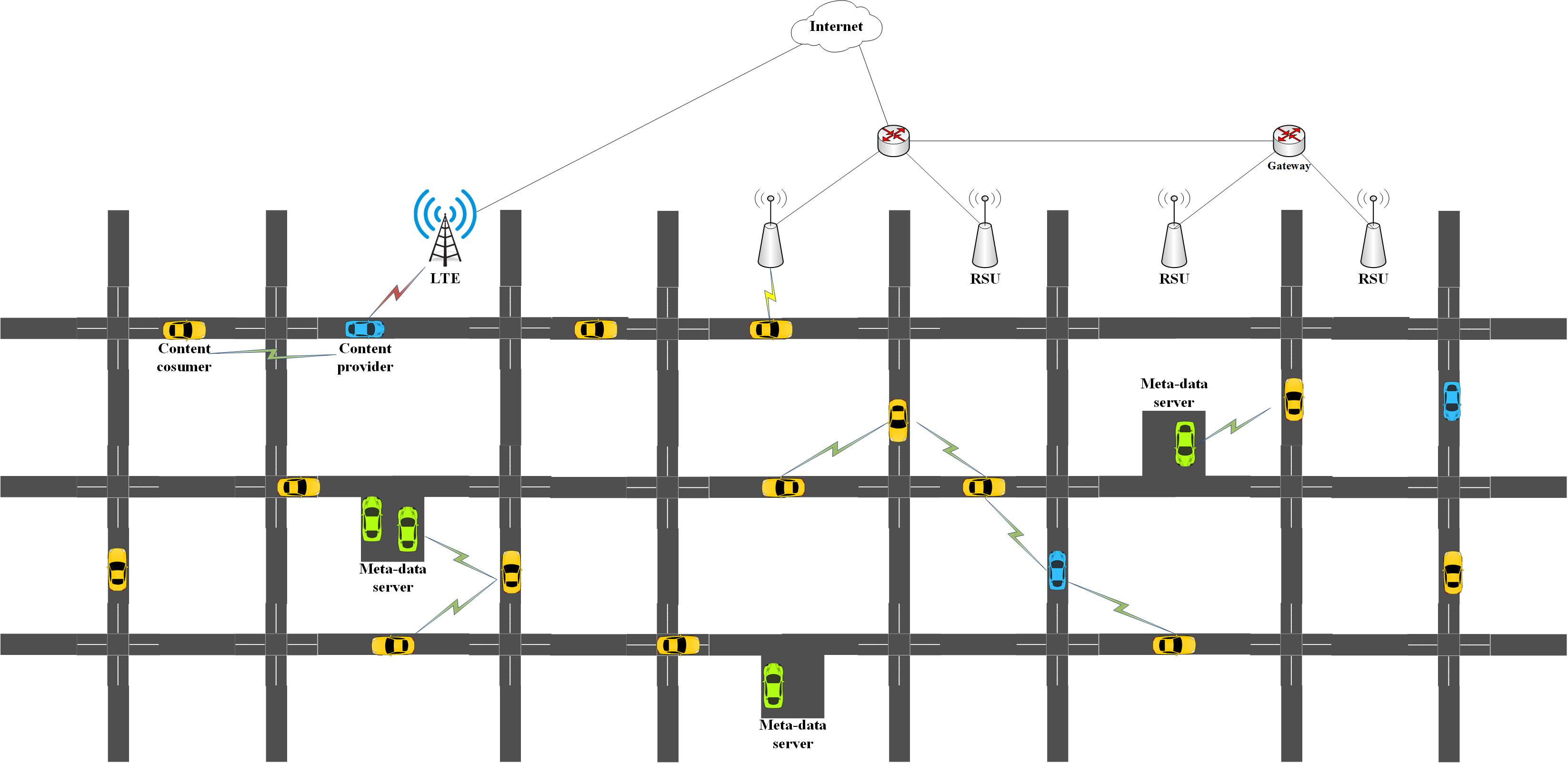}
	\caption{Content dissemination architecture}
	\label{architecture}
\end{figure*}

In vehicular networks, the content can be delivered through V2I connection with the RSU that is connected to the Internet, or though the cellular base station using 4G/5G interface \cite{Aung2018}. The former method has the advantage of being cheap and convenient communication method but difficult to access directly, and the vehicle need to rely on V2V communication to reach the sparse RSUs \cite{Aung2021}. While the latter has the advantage of wide coverage and instance access, but the drawback of expensive communication \cite{Dhelim2016}. To solve this problem, VeSoNet leverages a hybrid data distribution approach, where only the popular data content is replicated and store in data provider vehicles to minimize the download time and avoid excessive simultaneous downloads from 4G/5G networks. In this regard, we distinguish three types of vehicular nodes, (1) content consumer vehicles which form the majority of the vehicles in the network, and represent the end-user of the system. (2) content provider vehicles: vehicles that store the social network data, their objective is to maximize their revenue by delivering data content to content consumer vehicles as they travel through the city. (3) meta-data vehicles: vehicles that are situated in busy locations across the city, they provide information about content location, and perform various computational tasks, such as shortest social path calculation, content similarity calculation. For example, in Fig. \ref{architecture}. content consumer vehicles are colored in yellow, content provider vehicles are colored in blue and meta-data servers are colored in green. Note that meta-data servers are chosen at busy locations across the city such as parking lots, where vehicles are always present and not moving frequently which ensures the quality of content lookup service. Meta-data server vehicles maintain a content indexing table that contains a list of available contents in the network, and a list of content provider vehicles that stores these contents; as well as, traffic information about these data provider vehicles, such as their expected path, and the latest reported location. Content providers send frequent updates to meta-data severs to update their location and their expected travel path. To deliver the data content to data consumer vehicles, VeSoNet follows Information Centric Networking (ICN) model. When a data consumer vehicle is interested in a given content, it creates an interest packet regarding the desired content that contains the content identifier as well as the traffic information of the requesting vehicle, such as the expected travel path. The interest packet is broadcasted to all neighboring nodes and forwarded to other nodes until it reaches either a content provider node that stores this content, or meta-data vehicle that knows in which data provider vehicle this content is stored. When an intermediate node receives an interest packet and it does not stores the requested content, it forward the interest packet to the nearest meta-data vehicles. In the case where none of the data provider vehicles has the requested content, the content is downloaded from external network through RSU and forwarded back to the requester vehicle, and backed up in data provider vehicles for future requests.

\section{System modeling}

\subsection{Content consumers path planning}

\begin{figure}[!htbp]
	\centering
	\includegraphics[width=\columnwidth]{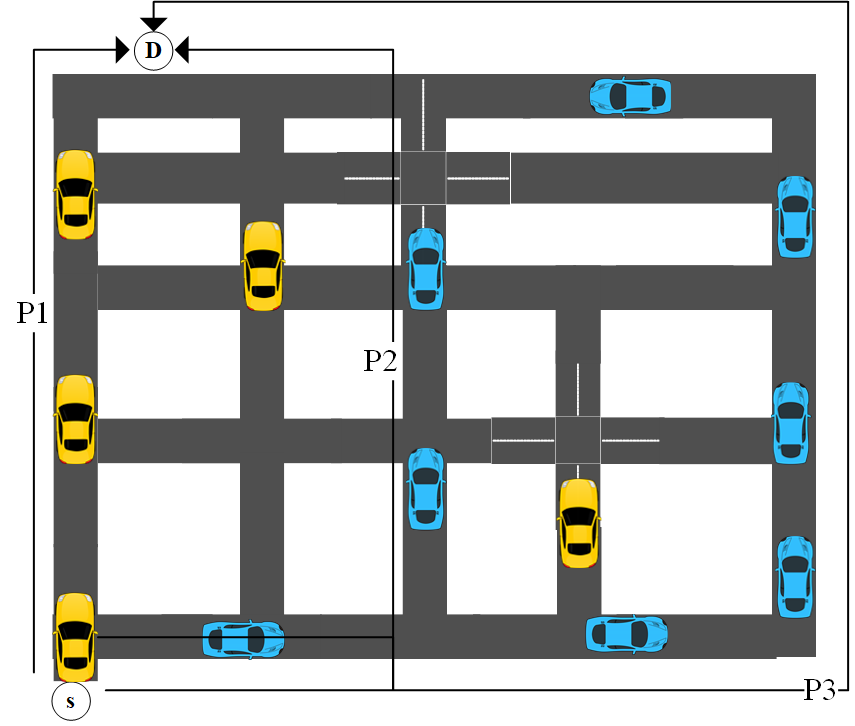}
	\caption{Social path selection}
	\label{social_path}
\end{figure}

The proposed framework leverages traffic information and change vehicles' traveling path to bring the content consumers close to content providers, hence enhancing content delivery experience. As a content provider takes the same path with content consumers, the delivery delay is significantly reduced. For instance in Fig. \ref{social_path}, a content consumer is traveling from the source location (S) to the destination location (D). Although the path P1 is the shortest path, P2 is recommended since it contains more content providers, and does not exceed the rerouting threshold as P3 does. Formally, let $P_{sh}=\{I_s\to I_{x1},\dots ,I_{xn}\to I_d\}$ be the shortest path computed based only on traffic information, without considering the availability of content providers, $I_s$ and $I_d$ are the starting intersection (source intersection) and destination intersection respectively. The objective is to compute the alternative social-aware shortest path $P_{so}$ that maximizes the number of content providers for the current vehicle during its trip, under the constrain that the difference between shortest travel time $\tau \left(P_{sh}\right)$ and social path $\tau \left(P_{so}\right)$ is less than the path optimization threshold $\varepsilon$. The na\"{i}ve approach is to recompute the shortest path between the starting $I_s$ and $I_d$ and consider the social path that maximizes the content providers and satisfy the path optimization threshold $\varepsilon $. However, this approach is computationally expensive, therefore, a graph pruning search algorithm that computes the social path as presented in Algorithm \ref{ShortestSocialPath}. The proposed algorithm is computationally efficient, as it leverages path pruning technique to eliminate the paths that exceed the path optimization threshold $\varepsilon $, hence minimize the search domain. The Algorithm takes the traffic-only shortest path and go through each road segment, by keeping track of the current processed intersection $n_{cur}$ (current node) and the next intersection $n_{next}$ (next node). In this context, $P_{temp}$ (temporary social path) is defined as the optimal social path between $n_{cur}$ and $n_{next}$ that satisfies the path optimization constrain. $P_{par}$ (partial social path) is defined as the optimal social path between the starting intersection $I_s$ (source node) and the next intersection $n_{next}$.  $P_{can}$ (candidate social path) is defined as the concatenation of currently known optimal social path from the starting intersection $I_s$ till the next intersection $n_{next}$ (note that this path is dynamically changing, as the alternative social-aware shortest path $P_{so}$  is updated over the time) and the temporary social path $P_{temp}$. After computing the afore-mentioned paths, if they newly found partial social path $P_{par}$ satisfies the path optimization threshold constrain ($\tau \left(P_{par}\right)-\tau \left(P_{can}\right)\le \varepsilon $), and the content providers in  $P_{par}$ is more than those in $P_{can}$, then $P_{so}$ is set as $P_{par}$. Otherwise, $P_{so}$ updated to include the newly found $P_{temp}$.

\begin{algorithm}
	\caption{SHORTEST\_SOCIAL\_PATH}
	\label{ShortestSocialPath}
	\hspace*{\algorithmicindent} \textbf{Input $P_{sh}=\{I_s\to I_{x1},\dots ,I_{xn}\to I_d\}$} \\
	\hspace*{\algorithmicindent} \textbf{Output  $P_{so}=\{I_s\to I_{y1},\dots ,I_{yn}\to I_d\}$} 
	\begin{algorithmic}[1]
		\STATE $P_{so}\leftarrow \{I_s\to I_{x1}\}$
		\STATE $n_{cur}\leftarrow I_s$
		\WHILE {$n_{cur}\neq I_d$}
		\STATE $n_{next}\leftarrow P_{sh}\left[next\left(n_{cur}\right)\right]$
		\STATE $P_{temp}\leftarrow AlternitiveSocialPath(n_{cur},n_{next})$
		\STATE $P_{par}\leftarrow AlternitiveSocialPath(I_s,n_{next})$
		\STATE $P_{can}\leftarrow P_{so}\left[I_s\leadsto n_{cur}\right]\cup P_{temp}$
		\IF {$\tau \left(P_{par}\right)-\tau \left(P_{can}\right)\le \varepsilon $}
		\IF {$\left|Providers\left(P_{can}\right)\right|<\left|Providers\left(P_{par}\right)\right|$}
		\STATE $P_{so}\leftarrow P_{par}$
		\ELSE
		\STATE $P_{so}\leftarrow P_{so}\cup P_{temp}$
		\ENDIF
		\ENDIF
		\STATE $n_{cur}\leftarrow P_{so}\left[Last\right]$
		\ENDWHILE
	\end{algorithmic} 
\end{algorithm}

The computation of alternative social path between two intersections is computed as shown in Algorithm \ref{AlternativeSocialPath} where it keeps track of the currently found optimal path  $P_{opt}$ and its travel time $t_{opt}$ and data providers ${so}_{opt}$, as well as a set that marks the previously visited nodes $VIST$. The core idea of the proposed graph pruning algorithm is to start from the starting node (intersection) and recursively evaluate the neighbors ($\mathop{n_{cur}}$) , and stop the evaluation once the current road segment proves to exceed the path optimization threshold $\varepsilon $ or the current processed node have been previously visited and we already know the rest of branches' outcome. In such way, the search space is considerably minimized, which reduce the computational cost of finding alternative social-aware path.

\begin{algorithm}
	\caption{ALTERNATIVE\_SOCIAL\_PATH}
	\label{AlternativeSocialPath}
	\hspace*{\algorithmicindent} \textbf{Input $s,d$} \\
	\hspace*{\algorithmicindent} \textbf{Output  $P_{opt}$} 
	\begin{algorithmic}[1]
		\STATE $P_{opt}\leftarrow P_{sh}\left[s\leadsto d\right]$
		\STATE $t_{opt}\leftarrow \tau (P_{sh}\left[s\leadsto d\right])$
		\STATE ${so}_{opt}\leftarrow \left|Providers\left(P_{sh}\left[s\leadsto d\right]\right)\right|$
	    \STATE $VIST\leftarrow \left\{(s,0)\right\}$
	    \STATE SOCIAL\_GRAPH\_PRUNING$\left(\left\{s\right\},d,0,0\right)$
	\end{algorithmic} 
\end{algorithm}

\begin{algorithm}
	\caption{SOCIAL\_GRAPH\_PRUNING}
	\label{SocialPathPurning}
	\hspace*{\algorithmicindent} \textbf{Input $P,d,t,so$} \\
	\hspace*{\algorithmicindent} \textbf{Output  $P_{opt}$} 
	\begin{algorithmic}[1]
		\STATE $n_{cur}\leftarrow P\left[Last\right]$
		\IF {$\left(n_{cur}\notin VIST\right)$ OR\newline
			 $\left({(n}_{cur}\in VIST\right)  AND  \left(VIST\left[n_{cur}.t\right]\le t\right)$ \newline AND $(VIST\left[n_{cur}.so\right]\ \ge so)$}
		\IF {$n_{cur}\neq d$}
		\FORALL {$n_{next}\in \mathop{n_{cur}}$}
		\STATE $t_{new}\leftarrow t+\ \tau (n_{cur}\leadsto n_{next})$
		\IF {$t_{new}\ \le \ \tau (P_{sh}\left[s\leadsto d\right])$}
		\STATE ${so}_{new}\leftarrow so+\ Providers(n_{cur}\leadsto n_{next})$
		\STATE $VIST\leftarrow VIST\cup \left\{n_{cur},t_{new},{so}_{new}\right\}$
	\STATE  SOCIAL\_GRAPH\_PRUNING\newline$\left(P\cup \left\{n_{cur}\leadsto n_{next}\right\},d,t_{new},{so}_{new}\right)$
		\ELSE
		\STATE $VIST\leftarrow VIST\cup \left\{n_{cur},t_{new},{so}_{new}\right\}$
		\STATE RETURN
		\ENDIF
		\ENDFOR
		\ELSE
		\IF {$t_{new}\ \le \ \tau (P_{sh}\left[s\leadsto d\right])$) and (${so}_{new}\ \le \ {so}_{opt}$}
		\STATE ${so}_{opt}\leftarrow {so}_{new}$
		\STATE $p_{opt}\leftarrow P$
		\ENDIF
		\ENDIF
		\ENDIF
	\end{algorithmic} 
\end{algorithm}

\subsection{Traffic-aware content recommendation scheme}

Given a data provider vehicular node $v_x$ that stores the set of social content items $c_x=\left\{c_{x1},c_{x2},\dots ,c_{xi}\right\}$, and is taking the path $p_x=\left\{i_{x1},i_{x2},\dots ,i_{xk}\right\}$, and the expected consumer vehicles $\mathrm{\Lambda }(v_x)$ when $v_x$ crosses $p_x$.  At each traffic light stop of the road intersections in $p_x$, $v_x$ tries to request the recommended content items stored in other nearby content providers. Formally, let $v_y$ be a nearby content providers that stores the set of social content items $c_y=\left\{c_{y1},c_{y2},\dots ,c_{yi}\right\}$. Given the previously viewed/liked content items of each consumer vehicle in $\mathrm{\Lambda }(v_x)$, which items in $c_y$  are most likely be viewed/liked by the consumer vehicles $\mathrm{\Lambda }(v_x)$. 

We formulate this situation as link prediction problem, where the objective is to predict the existing of potential links between the previously consumed content of expected consumer vehicles and the available content in nearby content provider vehicles. Since the waiting time in intersections is relatively short, therefore computing the potential links to recommended content for download using classical filtering recommendation methods, such as matrix factorization using collaborative filtering is not suitable in this situation. That is because such methods computational expensive and require knowledge about all the consumption history of all other vehicles in the systems to find the recommendation neighborhood.  Therefore, we propose a graph embeddings based content recommendation approach named vehicle2vec, where the vehicles are represented by a set of low dimensional vectors of the previously consumed content. vehicle2vec first learns the feature representations of each content available in the system, the content network is represented as graph data structure, where the nodes represent the data content and the edge represent the content similarity between these nodes. To learn the content nodes embeddings, vehicle2vec learns the content node low dimensional vector that preserves local neighborhoods of nodes in the original graph by solving an optimization problem \cite{grover2016node2vec}. In order to learn such content node embeddings, vehicle2vec leverages stochastic gradient descent (SGD) to optimize the objective function, hence learn the low dimensional representation.  Formally, given the content network represented as graph structure $G_c=\left(V_c,E_c\right)$, where $V_c$ are the data content nodes, and $E_c\subseteq \left(V_c\times V_c\right)$ is the set of edges links the content nodes. The objective of network embedding learning is to represent each content nodes ${c\in V}_c$ into a vector of low dimensional space $R^d$, which involve evaluating the mapping function $f_c:V_c\to R^d$, where $d\ll \left|V_c\right|$ is the size of feature in low-dimensional space where the original network structure is preserved. To learn the network embeddings, for each content node $c\in V_c$, we compute the content network neighborhood $N(c)\subset V_c$ that represent the semantically similar data content. We leverage the Skip-gram model of networks, hence the mapping function $f_c$ can be evaluated by optimizing the objective function presented in (\ref{maximize_equasion})

\begin{equation}
	\label{maximize_equasion}
	{\mathop{\mathrm{max}}_{f} \sum_{c\in V_c}{{\mathrm{log} \mathrm{Pr}\mathrm{}\left(N(c)\mathrel{\left|\vphantom{N(c) f_c(c)}\right.\kern-\nulldelimiterspace}f_c(c)\right)\ }}\ }
\end{equation}	
\[\] 

According to the symmetry property in the feature space, where the proximity between every pair of nodes is symmetric. Therefore, the conditional likelihood between every content node and its neighbors can be modelled as softmax unit where the parameter is the dot product of the nodes' feature vectors, as shown in (\ref{feature_vector_equasion})

\begin{equation}
	\label{feature_vector_equasion}
{\mathrm{Pr} \left(n_i\mathrel{\left|\vphantom{n_i f_c\left(c\right)}\right.\kern-\nulldelimiterspace}f_c\left(c\right)\right)\ }=\frac{\mathrm{exp}\mathrm{}(f_c\left(n_i\right)\cdot f_c\left(c\right))}{\sum_{m\in V_c}{\mathrm{exp}\mathrm{}(f_c\left(m\right)\cdot f_c\left(c\right))}}
\end{equation}	

In VeSoNet, every consumer vehicle is represented its vehicle2vec matrix that incorporate all the embedding feature of the previously consumed content. The content recommendation algorithm is described in Algorithm \ref{Intersection_recommendation}. where $v_x$ is the data provider that is taking the path $p_x$.In each intersection in $p_x$, $v_x$ checks the content available in every content provider vehicle $v_y$, and compute its similarity with the vehicle2vec low space features of all its expected content consumer vehicles $\mathrm{\Lambda }\left(v_x\right)$, if the similarity is greater than the content similarity threshold $\alpha $, then that content item is downloaded during the intersection waiting time.

\begin{algorithm}
	\caption{INTERSECTION\_RECOMMENDATION}
	\label{Intersection_recommendation}
	\hspace*{\algorithmicindent} \textbf{Input ${v_x,\ P}_x$} \\
	\hspace*{\algorithmicindent} \textbf{Output  $R_x$} 
	\begin{algorithmic}[1]
		\FORALL {$v_y\ in\ i_x$}
		\FORALL {$v_y\ in\ i_x$}
		\FORALL {$c_j\ in\ v_y$}
		\FORALL {$\mathrm{\ m}\in vehicle2vec(\mathrm{\Lambda }\left(v_x\right))$}
		\IF {$\left(\frac{\sum_{k\in \mathrm{m}}{sim\left(vec(k),{vec(c}_j)\right)}}{\left|\mathrm{m}\right|}\ >\alpha \right)$}
		\STATE $R_x\leftarrow R_x\cup \left\{c_j\right\}$
		\ENDIF
		\ENDFOR
		\ENDFOR
		\ENDFOR 
		\ENDFOR
	\end{algorithmic} 
\end{algorithm}

\subsection{Content provider distribution}

The problem of computing the minimum-cost combined sequence of routes serving a given number of customers is traditionally known as the vehicle routing problem (VRP), which is considered an NP-hard problem \cite{koh2020real}. Recently, various optimization algorithms have been proposed as an alternative where a sub-obtimal solution is computed under given constraints, such as firefly algorithm \cite{altabeeb2019improved}, genetic algorithm \cite{mohammed2017solving} or hybrid algorithm \cite{pan2021hybrid}. However, in these heuristic optimization algorithms compute the optimal solution under the assumption of stable traffic condition. Unlike the traditional VRP problem, in our situation the traffic flow is constantly changing over the time, and the search for solution is performed under dynamic traffic condition.  In our case, a given content provider vehicle $v_x$ that is traveling from starting position ${PoS}_s$ and traveling to ${PoS}_D$ through the path $p_x$, the objective of $v_x$ is to maximize its revenue that is generate from the advertisements delivered to the content consumers. The intuitive approach will be to choose road that maximize the number of content consumer vehicles, but if the same road contain large number of data provider vehicles, which will minimize the generated revenue. We define the content deliverability $CD\left(r_x,t_k\right)$ in road segment $r_x$ at time slot $t_k$ as the difference number of content consumer vehicles (CC) and content provider vehicles (CP), as shown in (\ref{cd_equasion}).

\begin{equation}
\label{cd_equasion}
CD\left(r_x,t_k\right)=CC\left(r_x,t_k\right)-CP\left(r_x,t_k\right)
\end{equation}

We proposed a social-aware DRL approach to optimize the content provider vehicles distribution. Reinforcement learning models evaluate the action taken in each step in the system that change the states, and through reward and punishment; the agent learns to  maximize the performance. In the proposed social-aware DRL model, the states represent the current road segment content deliverability. While the actions represent the navigation decision taken by content provider vehicles in road intersections during its travel path. And as the content provider vehicles' objective is to maximize the content deliverability, the reward is computed as the difference of the old and new content deliverability value after taken that action, as shown in (\ref{r_equasion}). 
\begin{equation}
	\label{r_equasion}
	r=O_{cd}-N_{cd}
\end{equation}

The proposed model is trained using Deep Q-Network (DQN) approach, in which we leverage Q-learning system that is used to choose the optimal option for all state/action pairs by evaluation the Q value of that action, denoted as $Q\left(s_n,a_s\right)$ as shown in (\ref{q_equasion}), where $s_o$ represent the previous state, $a_o$ is the previously taken action, and $r$ denotes the obtained reward by taken that action, and $\gamma $ denotes the discount factor. 
\begin{equation}
	\label{q_equasion}
	Q\left(s_n,a_s\right)=Q\left(s_o,a_o\right)+\beta \left(r+\gamma \ Q\left(s_n,a_s\right)\right)-Q\left(s_0,a_0\right)
\end{equation}

The input is fed into multilayer neural network that is used as function that maps the system state to the corresponding Q-value. After the training, the network is used to predict the optimal action to take in a given situation. At each intersection, in order to maximize the content deliverability reward $r$. As the content data provider is aiming to take the optimal action in current state s using the pre-trained network, as presented in (\ref{q2_equasion}), where ${max}_{\grave{a}}Q(s,a)$ is the list Q-values yielded from the neural network output; Therefore, we choose the action that yielded the maximum rewards, as shown in (\ref{a_equasion}) where $\ s,a$ are the current action and state.
\begin{equation}
	\label{q2_equasion}
	Q\left(s_n,a_s\right)\leftarrow r\left(s_n,a_s\right)+\gamma \ {max}_{\grave{a}}Q(s,a)
\end{equation}
\begin{equation}
	\label{a_equasion}
	a=arg{max}_{\grave{a}}\ Q(s,a)
\end{equation}

All the training data is memorized, as it is stored replay buffer. Content provider vehicles explore the environment and take random actions and also by looking up the suggested value from Q-network, which is known as epsilon-greedy policy. The loss function is utilized to compute the squared difference between the predicted and actual values, hence the loss is minimized by updating the network weights, as shown in (\ref{L_equasion}).

\begin{equation}
	\label{L_equasion}
	L={(r+\gamma \ {max}_{\grave{a}}Q\left(s_n,a_n;{\theta }_n\right)-Q(s_0,a_0;{\theta }_0))}^2
\end{equation}

\section{Experimental evaluation}

\subsection{Experiment and dataset}

To simulated the proposed system, we have used the simulator of urban mobility (SUMO) \cite{behrisch2011sumo} , version 0.30.0. For the network simulation, we have used OMNeT++ network simulator \cite{varga2010omnet}, version 5.0, and used the vehicular network framework Veins \cite{sommer2019veins}, version 4.6 on top of OMNeT++. We have used a real map of Greater London to simulate the traffic, the map raw data were extracted from OpenStreetMap \cite{osm}, the map was further processed by Netconvert tool to generate the road network that is used by SUMO to generate the traffic. Table \ref{simulation_parameters} shows a detailed description of the network simulation parameters. We have used Last.fm dataset \cite{GroupLens} to simulate the social network content. The dataset contains the social networking and music listening information of more than 2k users. Vehicles are randomly placed in the map, and they travel to random destination location. Each vehicle is randomly associated with two users from Last.fm dataset, and we consider them as the passengers within that vehicle. Once the vehicle is added to the simulation, we first measure the shortest path to its destination point, and estimate the required travel time, after that the vehicle is routed according to the shortest social path according to our proposed algorithm. 

\begin{table}
	\centering
	\caption{Simulation parameters}
	\label{simulation_parameters}
	\begin{tabular}{|l|l|l}
		\cline{1-2}
		\textbf{Parameter}   & \textbf{Value}   \\ \cline{1-2}
		PHY model            & IEEE 801.11p     \\ \cline{1-2}
		MAC model            & EDCA             \\ \cline{1-2}
		Propagation model    & Two rays         \\ \cline{1-2}
		Fading model         & Rayleigh fading  \\ \cline{1-2}
		Antenna model        & Omnidirectional  \\ \cline{1-2}
		Shadowing model      & LogNormal        \\ \cline{1-2}
		Channel frequency    & 5.890e9 Hz       \\ \cline{1-2}
		Propagation distance & 450m             \\ \cline{1-2}
		Transmission power   & 20 mW            \\ \cline{1-2}
		Maximum hop count    & 15               \\ \cline{1-2}
		Scenario map         & London           \\ \cline{1-2}
		Scenario area        & $5km^2$          \\ \cline{1-2}
	\end{tabular}
\end{table}

\subsection{Baselines and metrics}

To evaluate the performance of VeSoNet, we have compared it with the following systems described in related works: Cuckoo \cite{alowish2020novel}, Safe and Sound (SNS) \cite{de2019safe}, CCN-CF \cite{soua2017sdn}, and DP-IB \cite{zhao2007data}. The proposed system VeSoNet is compared to the afore-mentioned baselines using various metrics, the performance of each system will be assessed by the following evaluation metrics:

Delivery delay: The time required to deliver the requested content, this metric is computed by measuring the time difference between the timestamp when the content consumer solicited the content and the timestamp when it actually received the content. 

Delivery rate: This metric is computed as a complementary of the delivery delay metric, since measuring the undelivered content delay will considerably increase the delay, hence prevent the accurate measurement of successfully delivered content. For this reason, the delivery rate is computed as a separate measure that represent the ratio of successfully delivered content from the total requested content.

Travel time: The required time to travel from the starting location to the destination location. This metric is used to measure the additional travel time eventuated by taking alternative social path rather than the shortest travel path.

Computation cost: This metric measures the number of computational operations required to perform the computational task of the system, such as path rerouting, content recommendation and traffic prediction. 

\subsection{Results analysis}

\begin{figure*}[!htbp]
	\centering
	\includegraphics[width=\textwidth]{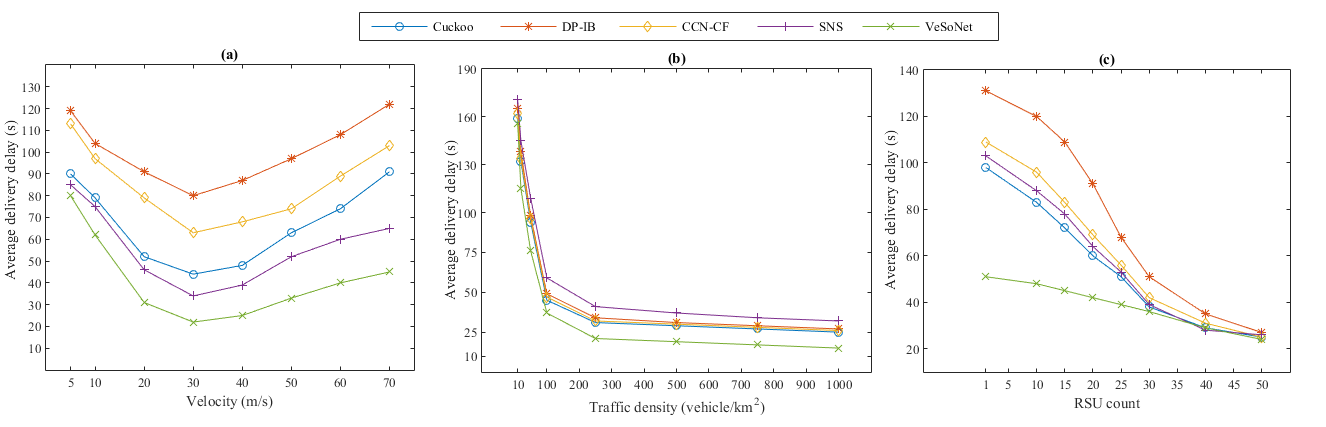}
	\caption{Comparing average delivery delay under (a) increasing velocity (b) increasing traffic density (c) different RSUs count}
	\label{result_A}
\end{figure*}

Fig. \ref{result_A} shows the average delivery delay of the studies systems in different settings. Where in Fig. \ref{result_A} (a), we increase the velocity of the vehicles and observe the delivery delay, and in Fig. \ref{result_A} (b), the traffic density is increase to measure its impact on the delivery delay, and Fig. \ref{result_A} (c) shows the average delivery delay values with different number of RSUs.  In Fig. \ref{result_A} (a), we can see that all the compared systems have relatively high delivery delay in low velocity environments (less than 20m/s). That is because in low velocity environments, the content requests, as well as the content delivery messages that rely on V2V are slowed down in store-and-carry situation, when there is no nearby vehicles. The moderate velocity settings (25m/s to 45 m/s) yield the shortest delivery delay for all systems. While in high velocity settings (more than 45 m/s), the delivery delay increases again. That comes as a result of the frequent disconnections due to the high velocity of the vehicles, which makes difficult to establish V2V communication, hence increase the delivery time of content request and reply as well. In Fig. \ref{result_A} (a), we can observe that the VeSoNet and other vehicle rerouting and path planning schemes (SNS and Cuckoo) clearly outperform the CCN-CF that does not leverage path planning or vehicle rerouting, and also outperform DP-IB, which relies on the A-Roads and C-Roads intersection to deliver the content. That is because, in VeSoNet, the vehicle rerouting strategy brings the content provider and the content consumer vehicles near to each other's, which eliminate the need for intermediate vehicles, SNS and Cuckoo also employ similar rerouting strategies. Fig. \ref{result_A} (b) shows the average delivery delay in different vehicle density environments. In which we can observe that all the studied systems have longer delivery delay in low-density environments (10-50 vehicles), that is because in such sparse environments, it become extremely difficult to forward messages between vehicles, where the vehicles need to travel long distance to find other forwarder vehicles. However, the deliver delay is considerably shortened in high-density environments (more than 100 vehicles).  VeSoNet still have the upper hand compared to other baselines, that is due to the social-aware rerouting strategy and also popular content replication method, as the density increases, the demand for such popular content increases accordingly, which contribute to the decrease of average delivery delay.  Fig. \ref{result_A} (c) shows the average delivery delay with different number of deployed RSUs. The objective of this test is to measure the ability of the studied systems to cope with network failures and their ability to work in infrastructure-less environments. With high RSUs count, all the baselines have similar delivery delay, however, as the RSUs count is reduced, we can clearly see the superiority of VeSoNet, that is due to its distributed nature. Where most of the popular contents are stored in content provider vehicles and distributed across the network and indexed in meta-data servers, where the content consumer can request and get the content through V2V without the need for V2I communication, unless the requested content is downloaded for the first time. DB-IP has the worst performance among other baselines, as it mainly depends on RSU to broadcast the content in A-Roads, and as the number of RSU is reduced, the number of A-Roads is reduced as well, hence the intersections between A-Roads and C-Roads is minimized as well.

\begin{figure*}[!htbp]
	\centering
	\includegraphics[width=\textwidth]{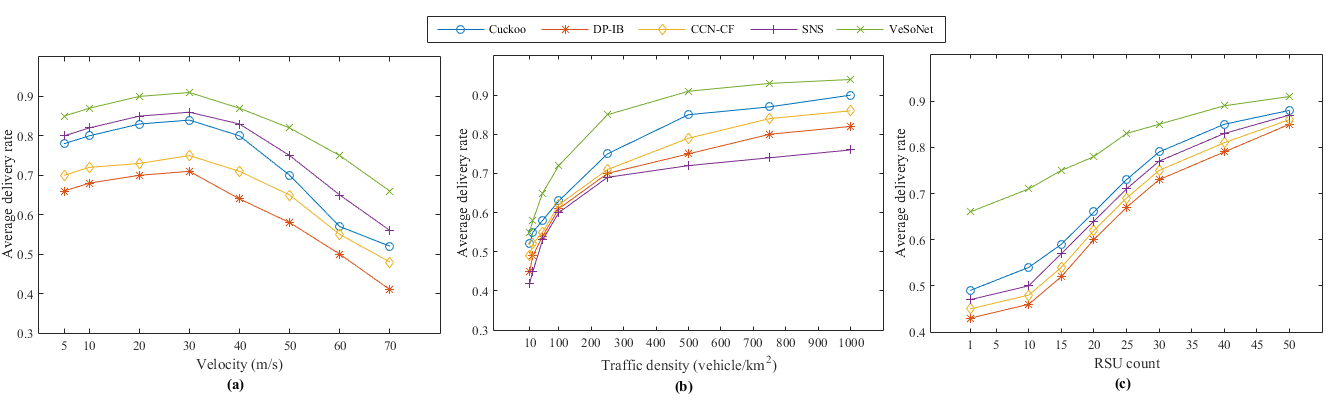}
	\caption{Comparing average delivery rate under (a) increasing velocity (b) increasing traffic density (c) different RSUs count}
	\label{result_B}
\end{figure*}

Fig. \ref{result_B}  shows the average delivery rate of the studied baselines in different velocity, density and infrastructure connectivity environments. Where in Fig. \ref{result_B} (a), we increase the velocity of the vehicles and observe the delivery rate, and in Fig. \ref{result_B} (b), the density of the vehicles is increase to measure its impact on the delivery rate, and Fig. \ref{result_B} (c) shows the average delivery rate with various RSUs count. In Fig. \ref{result_B} (a), we can observe that all the studied systems have high delivery rate in low and medium velocity environments, but low delivery rate in high velocity settings. That is because of the frequent V2V communication disconnections in high velocity of the vehicles, which makes it difficult to establish communication and transfer the content, hence reduce the delivery rate. Similar to the delivery delay scenario, VeSoNet and other vehicle rerouting and path planning schemes (SNS and Cuckoo) still have the upper hand compared to other baselines, as the path rerouting allows to bring the content provider closer to the content requester which reduce the need for multiple-hop communications that do not perform well in high velocity settings. While in Fig. \ref{result_B} (b), we can observe that all the studied baselines have relatively low delivery rate in low density environments (less than 100 vehicles). That is because in such sparse environments, it becomes extremely difficult to forward messages between vehicles, where the vehicles need to travel long distance to find other forwarder vehicles, which may lead to the content interest timer expire and the delivery is considered failed. While in high-density environments, the delivery rate of all the baselines considerably increase, as V2V communication becomes easier in such dense settings. Similarly, in Fig. \ref{result_B} (c), we can see that the average deliver rate plummets as the we reduce the number of RSUs. Again, we can clearly observe the superiority of VeSoNet due to its distributed nature. Where most of the popular contents are stored in content provider vehicles and distributed across the network and indexed in meta-data servers, where the content consumer can request and get the content through V2V without the need for V2I communication, unless the requested content is downloaded for the first time. On the other extreme, DB-IP has the worst performance among other baselines, as it mainly depends on RSU to broadcast the content in A-Roads and C-Roads.

\begin{figure*}[!htbp]
	\centering
	\includegraphics[width=\textwidth]{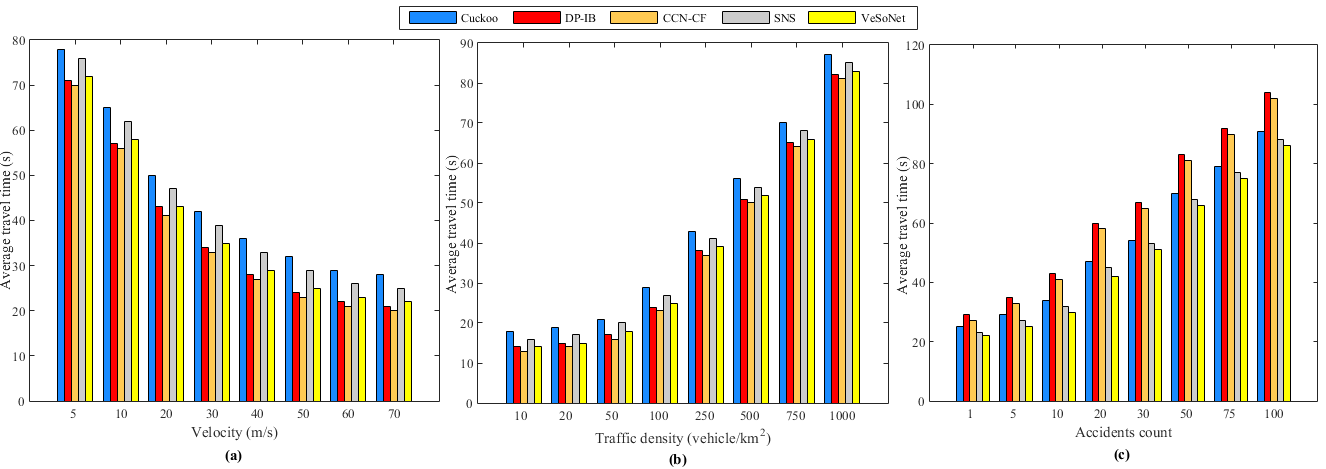}
	\caption{Comparing average vehicle travel time under (a) increasing velocity (b) increasing traffic density (c) increasing traffic accidents count}
	\label{result_C}
\end{figure*}

\begin{figure*}[!htbp]
	\centering
	\includegraphics[width=\textwidth]{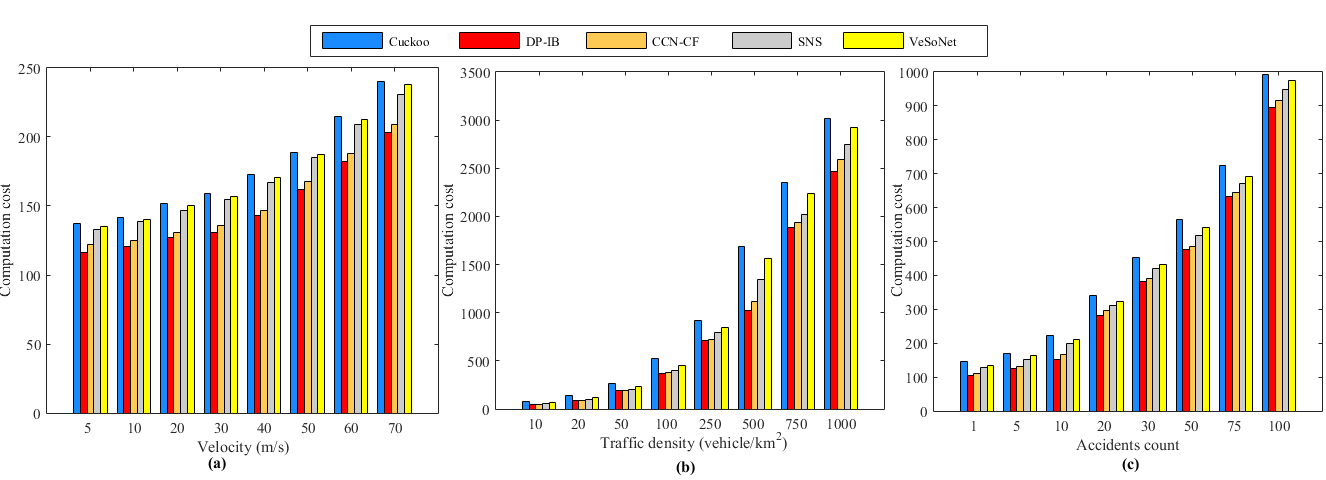}
	\caption{Comparing average system computational cost under (a) increasing velocity (b) increasing traffic density (c) increasing traffic accidents count}
	\label{result_D}
\end{figure*}

However, the extraordinary effectiveness of VeSoNet and other path planning path baseline in delivery delay and delivery rate comes with the cost of longer travel time and computational overhead, as shown in Fig. \ref{result_C} and Fig. \ref{result_D}. Fig. \ref{result_C} shows the average travel time of the compared systems in different velocity, density and accidents count settings. In Fig. \ref{result_C} (a), we increase the velocity of the vehicles and measure the travel time, and in Fig. \ref{result_C} (b), the density of the vehicles is increase to measure its impact on the travel time, and Fig. \ref{result_C} (c) shows the travel time values with increasing accidents count. From Fig. \ref{result_C} (a), we can observe that all the compared systems have longer travel time in low-velocity but settings (less than 30 m/s), but the travel time is exponentially decrease as we increase the maximum speed from 5 to 20. The travel time stabilize after 30, even the speed still increasing until 70, that is due to the waiting stops in the intersections, which do not allow the vehicles to reach their maximum speed, and even if they do, the quickly reach the next road intersection and have to start from 0 again. VeSoNet and other vehicle rerouting and path planning schemes (SNS and Cuckoo) have slightly longer travel time, which is because the re-routed paths are usually longer than the shortest travel paths. For example in VeSoNet the content consumer vehicles choose the path that have more relevant content providers, regardless of the shortest or quickest path, as long as the chosen path do not exceed the social shortest path threshold. Similarly, in Fig. \ref{result_C} (b), we can observe that VeSoNet and other vehicle rerouting and path planning schemes still have slightly longer travel time in all traffic density scenarios. However, in Fig. \ref{result_C} (c), the vehicle rerouting baselines have shorter travel time when the number of accidents increase, that is due to rerouting strategy that allows VeSoNet to avoid the accident roads.

Fig. \ref{result_D} shows the baselines' average computational in increasing content request count, density and accidents count settings. In Fig. \ref{result_D} (a), we increase the average content requests of the vehicles and measure the computational cost, and in Fig. \ref{result_D} (b), the density of the vehicles is increase to measure its impact on the computational cost, and Fig. \ref{result_D} (c) shows the computational cost values with increasing accidents count. In Fig. \ref{result_D} (a), we can observe that VeSoNet has relatively high computational cost than other baselines, that is because VeSoNet require the operation of multiple vehicular server that resolve the content requests, and more computational power is required to update and coordinate these vehicular server. However, in other baselines, the computation is concentrated in a centralized server, while in VeSoNet, the computation is distributed in various distributed server, and the computational cost in each server is considerably low. Similarly, in Fig. \ref{result_D} (b), the computational cost of all the studied systems exponentially increase with the increase of traffic density, and VeSoNet has the second highest computational cost, but keep in mind that this computational is distributed among various different vehicular servers, unlike other baselines. In Fig. \ref{result_D} (c), as the traffic accidents count increase, the computational cost of all the baselines increase, furthermore vehicle rerouting schemes have relatively higher computational cost due to the computational overhead yielded by re-computing the alternative travel paths.

\section{Conclusion}

In this paper, we have presented a traffic-aware vehicular content caching architecture that optimize the content dissemination among vehicles using social-aware graph pruning search algorithm that computes and assigns the vehicles to the shortest path with the most relevant vehicular content providers. To recommend relevant content according to their social context, a traffic-aware content recommendation approach based on graph embeddings is proposed, where the vehicles are represented by a set of low dimensional vector (vehicle2vec) of their previously consumed content. Experimental results show that the proposed architecture can reduce content delivery delay and delivery ratio more than 20\% compared to the state-of-the-art baselines, at the cost of slight increase in computational cost and average travel time. However, there are many aspects of future improvement:

\begin{itemize}
\item Although that all the communications between content consumers and content providers are encrypted, however, it is still possible to perform statistical attacks to infer the content consumers future paths, the privacy of the content consumers can be preserved by adding a pseudonyms identification scheme.

\item The vehicular edge computing architecture can be further extended by adding computational task offloading, where all the computational tasks are performed in the vehicles.

\item The social path selection process could be further extended to include driver preferences for individual road selection.

\end{itemize}

\section*{Acknowledgment}
This work was supported by the National Natural Science Foundation of China under Grant 61872038.

\ifCLASSOPTIONcaptionsoff
  \newpage
\fi



\bibliographystyle{IEEEtran}

\bibliography{refs}

%

%
\vskip 0pt plus -1fil
\begin{IEEEbiography}[{\includegraphics[width=1in,height=1.25in,clip,keepaspectratio]{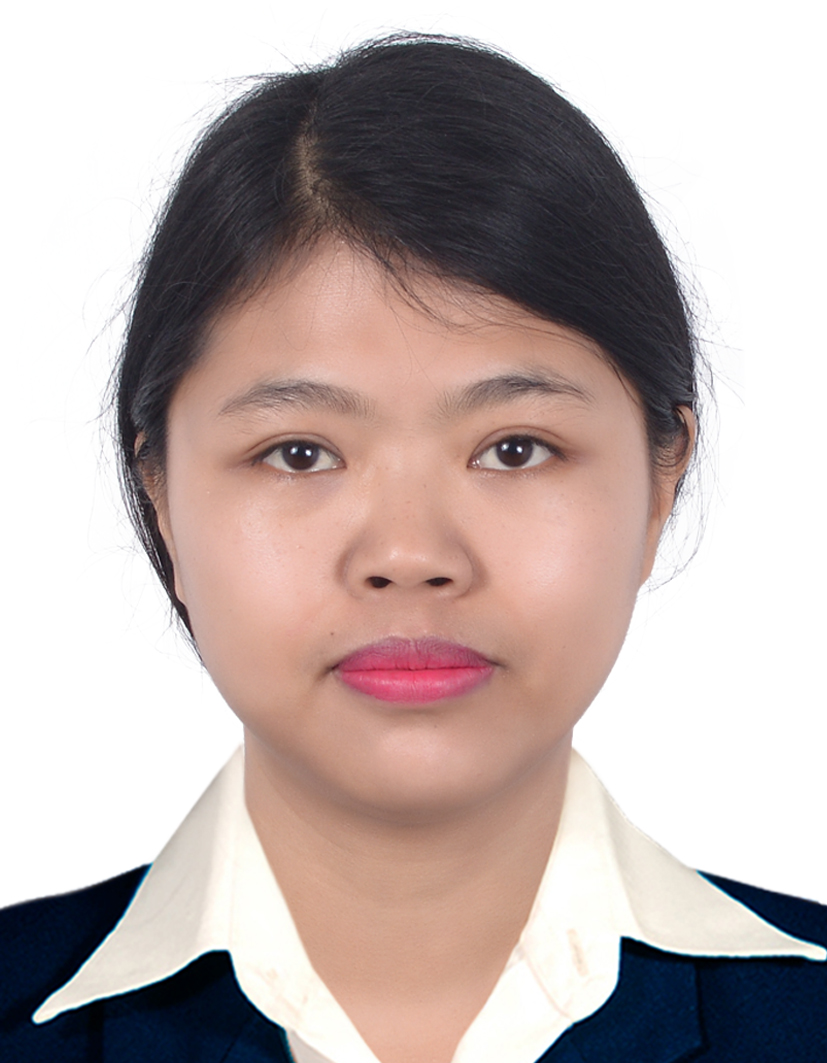}}]{Nyothiri Aung}
	is an assistant professor in Linyi University, China. She received Master of Engineering (Information Technology) degree form Mandalay Technological University, Myanmar, 2012. And PhD in Computer Science and Technology from University of Science and Technology Beijing, China, 2020. She worked as a tutors at the Department of Information Technology in Technological University of Meiktila, Myanmar (2008-2010). And System Analyst of ACE Data System, Myanmar (2012-2015). Her research interests include Social Computing, Personality Computing and Intelligent Transportation System.
\end{IEEEbiography}	

\vskip 0pt plus -1fil
\begin{IEEEbiography}[{\includegraphics[width=1in,height=1.25in,clip,keepaspectratio]{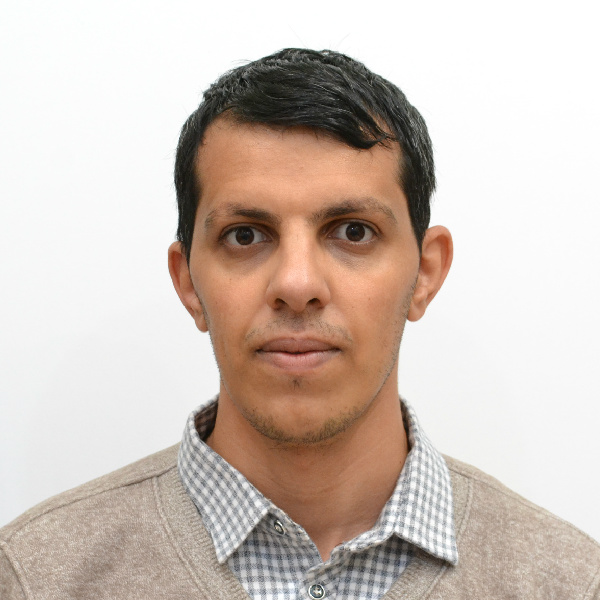}}]{Sahraoui Dhelim}
	is a postdoctoral researcher in University College Dublin, Ireland. He was a visiting researcher in Ulster University, UK (2020-2021). He obtained his PhD degree in Computer Science and Technology from University of Science and Technology Beijing, China, in 2020. And a Masters degree in Networking and Distributed Systems from the University of Laghouat, Algeria, in 2014. He serves as workshop chair of Cyberspace congress (CyberCon). His research interests include Social Computing, User Modeling, Deep-learning, Recommendation Systems and Intelligent Transportation Systems.
\end{IEEEbiography}

\vskip 0pt plus -1fil

\begin{IEEEbiography}[{\includegraphics[width=1in,height=1.25in,clip,keepaspectratio]{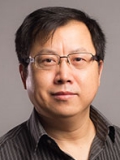}}]{Liming Chen}
	is a professor in the School of Computer  Science  and  Informatics  at  University  of  Ulster,  Newtownabbey,  United  Kingdom.  He  received his  B.Eng  and  M.Eng  from  Beijing  Institute  of Technology  (BIT),  Beijing,  China,  and  his  Ph.D  in Artificial Intelligence from De Montfort University,UK.  His  research  interests  include  data  analysis,ubiquitous computing, and human-computer interaction. Liming is a Fellow of IET, a Senior Member of IEEE, a Member of the IEEE Computational Intelligence Society (IEEE CIS), a Member of the IEEE CIS Smart World Technical Committee (SWTC), and the Founding Chair of the IEEE CIS SWTC Task Force on User-centred Smart Systems (TF-UCSS). He has served as an expert assessor, panel member and evaluator for UK EPSRC (Engineering and Physical Sciences Research Council, member of the Peer Review College), ESRC (Economic and Social Science Research Council), European Commission Horizon 2020 Research Program, Danish Agency for Science and Higher Education, Denmark, Canada Foundation for Innovation (CFI), Canada, Chilean National Science and Technology Commission (CONICYT), Chile, and NWO (The Netherlands Organisation for Scientific Research), Netherlands.
\end{IEEEbiography}

\vskip 0pt plus -1fil

\begin{IEEEbiography}[{\includegraphics[width=1in,height=1.25in,clip,keepaspectratio]{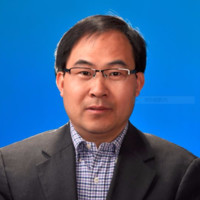}}]{Wenyin Zhang}
	is the full professor and the dean of the school of Information Science and Engineering, Linyi University, China. He received his MS from Shandong University of Science and Technology, and obtained his PhD in Computer Science from Chengdu Branch of the Chinese Academy of Sciences (CAS). He is the Vice Chairman of Shandong Computer Society, and the Dean of Linyi Smart Big Data Research Institute.  He is visiting scholar at the University of Southern Mississippi and Portland State University. His main research interests include computer vision, digital watermarking, network information security, blockchain technology application.  
\end{IEEEbiography}

\vskip 0pt plus -1fil

\begin{IEEEbiography}[{\includegraphics[width=1in,height=1.25in,clip,keepaspectratio]{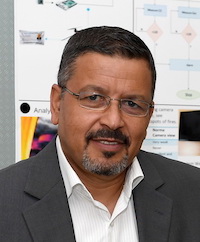}}]{Abderrahmane Lakas}
is a Professor at the Computer and Network Engineering department in the College of IT at UAE University. He holds an MS, and PhD in Computer Systems from the University of Pierre et Marie Curie (Paris VI, France). He has several years of both academic and industrial experience. He spent two years as a postdoc researcher at School of Computing and Communication at the University of Lancaster in UK. He is the head of CAST (Connected Intelligent Autonomous Systems) research group and the Connected Autonomous Intelligent Systems Lab (ASIL). Prior to joining UAE University, he held several industrial positions in several companies in Canada and the US including at Netrake (Plano, TX, USA), Nortel Networks (Ottawa, Canada), and Newbridge (Ottawa, Canada). His current research interests include intelligent transportation systems, vehicular ad hoc networks, unmanned ground and aerial vehicles, autonomous systems, smart cities, Internet of Things, and QoS. Dr. Lakas has published several research papers in scholarly journals. He is member of the TPC and reviewer of several renown conferences and serves in the editorial board of few journals.
\end{IEEEbiography}

\vskip 0pt plus -1fil
\begin{IEEEbiography}[{\includegraphics[width=1in,height=1.25in,clip,keepaspectratio]{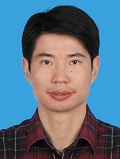}}]{Huansheng Ning}
	Received his B.S. degree from Anhui University in 1996 and his Ph.D. degree from Beihang University in 2001. Now, he is a professor and vice dean of the School of Computer and Communication Engineering, University of Science and Technology Beijing, China. His current research focuses on the Internet of Things and general cyberspace.
	He is the founder and chair of the Cyberspace and Cybermatics International Science and Technology Cooperation Base.
	He has presided many research projects including Natural Science Foundation of China, National High Technology Research and Development Program of China (863 Project). He has published more than 150 journal/conference papers, and authored 5 books. He serves as an associate editor of IEEE Systems Journal (2013-Now), IEEE Internet of Things Journal (2014-2018), and as steering committee member of IEEE Internet of Things Journal (2016-Now).
\end{IEEEbiography}





\end{document}